\documentclass[12pt]{article}

\newcommand{\epq}{\mbox{q}}
\newcommand{\epqa}{\overline{\mbox{q}}}

\newcommand{\bibl}[5]
	{#1, {\it #2} {{\bf #3},} #5 (#4)}
\newcommand{\cebe}{\begin{center}}
\newcommand{\ceen}{\end{center}}
\newcommand{\debe}{\begin{description} \vspace{-2ex}}
\newcommand{\deen}{\end{description}}
\newcommand{\eabe}{\begin{eqnarray}}
\newcommand{\eaen}{\end{eqnarray}}

\newcommand{\eqbe}{\begin{equation}}
\newcommand{\eqen}{\end{equation}}

\newcommand{\itbe}{\begin{itemize}}
\newcommand{\iten}{\end{itemize}}

\newcommand{\tabe}{\begin{tabbing}}
\newcommand{\taen}{\end{tabbing}}

\input{psfig}

\begin{document}

\begin{titlepage}
\begin{flushright}
  LU TP 97-07 \\
  April 1997
\end{flushright}
\vspace{25mm}
\begin{center}
  \Large
  {\bf Bose-Einstein Correlations in the Lund Model} \\
  \normalsize
  \vspace{12mm}
  Bo Andersson and  
  Markus Ringn\'{e}r\footnote{E-mail: bo@thep.lu.se, markus@thep.lu.se } 
  \vspace{1ex} \\
  Department of Theoretical Physics, Lund University, \\
  S\"olvegatan 14A, S-223 62 Lund, Sweden \\
\end{center}
\vspace{20mm}

\noindent {\bf Abstract:} \\
By providing the Lund Model fragmentation process with a quantum-mechanical 
framework we extend the results of \cite{ah86} to situations where there
are very many identical bosons. We investigate the features of the weight 
distributions in some detail and in particular exhibit three-particle 
BE-correlations, the influence on the $\rho$-spectrum and the difference 
between charged and neutral pion correlations. 
\vspace{3cm}



\end{titlepage}

\section{Introduction} \vspace{-2ex}
The Hanbury-Brown-Twiss (HBT) effect \cite{hbt56} 
originated in astronomy 
where one uses the interference pattern
of the photons to learn about the size of the photon emission region, i.e. the
size of the particular star, which is emitting the light. The effect can 
be described as an enhancement of the
two-particle correlation function that occur when the two particles 
are identical bosons and have very similar energy-momenta. A
well-known formula \cite{mgb85} to relate the two-particle 
correlation function (in four-momenta $p_j, j=1,2$ with $q=p_1-p_2$) 
to the space--time density distribution, $\rho$, of (chaotic) emission sources is, 
\eqbe
\frac{\sigma d\sigma_{12}}{d\sigma_1 d\sigma_2} = 1 + |{\cal R}(q)|^2
\label{e:ordinaryR}
\eqen 
where ${\cal R}$ is the normalised Fourier transform of
the source density  
\eqbe 
\label{calRform} 
{\cal R}(q) = \frac{\int \rho(x) dx \exp (iqx)}{\int \rho(x) dx} 
\eqen
This quantity is often, without very convincing reasons, parametrised 
in terms of a ``source radius'' $R$ and a ``chaoticity 
parameter'' $\lambda$, 
\eqbe
{\cal R}(q)= \lambda \exp(-R^2 Q^2)
\eqen
with $Q^2=-q^2$.
The source radii obtained by this parametrisation tend to be similar in all
hadronic interations (we exclude heavy ion
interactions where the extensions of nuclear targets and 
probes will influence the
result), with $R \sim 0.5-1~fm$, but the chaoticity varies rather much depending
upon the particular data sample and the method of the fit.
At present the knowledge of higher-order correlations is still limited in the
experimental data, although in principle there should be such correlations.

The HBT effect, being of an intrinsically quantum mechanical nature, is not
easily included in the event generator programs used in high energy physics.
Such simulation programs, like HERWIG \cite{herwig59} (based upon the 
Webber--Marchesini parton
cascades and ending by cluster fragmentation) and JETSET \cite{ts94} 
(based upon the Lund Model string dynamics \cite{ba83}) are built in 
accordance with classical stochastical processes, i.e. they produce a
probability weight  for an event without  any 
quantum mechanical interference effects.

Sj\"ostrand has introduced a clever 
device as a subroutine to JETSET, in which
the HBT effect is simulated as a mean field potential attraction between
identical bosons \cite{ls95}. Thus, given a set of energy-momentum vectors 
of identical bosons, $p_1,\ldots,p_n$, generated without any HBT effect, 
it is possible to
reshuffle the set into another set where each pair on the average has been
moved relatively closer to show a (chosen) HBT-correlation, while still keeping
to energy-momentum conservation for the whole event.

In this paper we will develop a method deviced in \cite{ah86} to provide the
Lund Model with a quantum mechanical interpretation. In particular there will
be a production 
matrix element with well-defined phases. This will then be used to make a model
of the HBT effect. Although this model stems from different considerations it
will 
nevertheless contain predictions which are similar to those in the
ordinary approach giving Eq~(\ref{e:ordinaryR}).

In the first section we survey those
features of the Lund model, that are necessary for the following. In the second
section we will exhibit the general $n$-particle HBT-correlations in the model.
The resulting expressions contain a sum of in general $n!$ terms, i.e. it is of
exponential type from a computational point of view. It is
possible to subdivide the expressions in accordance with the group structure of
the permutation group. Although the higher order terms provide small
contributions in general the computing times are still forbidding. 
In order to speed up the calculations we introduce instead the
notion of \emph{links} between the particles and it is in this way
possible to obtain expressions of power type from a calculational point of
view, which are perfectly tractable
in a computer. In the last section we exhibit a set of results both in order to
show the workings of the model and to provide predictions for experiments.

We will in this paper be satisfied to treat only two-jet events, i.e. we will
neglect hard gluon radiation and we will come back to HBT effects in gluon
events in another publication.

\section{Some Properties of the Lund Model} \vspace{-2ex}
Within the framework of perturbative QCD it is possible to obtain many useful
formulas but all the results are expressed in a partonic language. In order 
to be able to compare to the hadronic distributions, which are observed in the 
experimental setups, it is
necessary to supplement the perturbative results with a fragmentation process.
We will in this paper be concerned with the Lund string model \cite{ba83} 
and we start with a brief introduction to its main properties.

\begin{figure}[t]
  \hbox{\vbox{
    \begin{center}
    \mbox{\psfig{figure=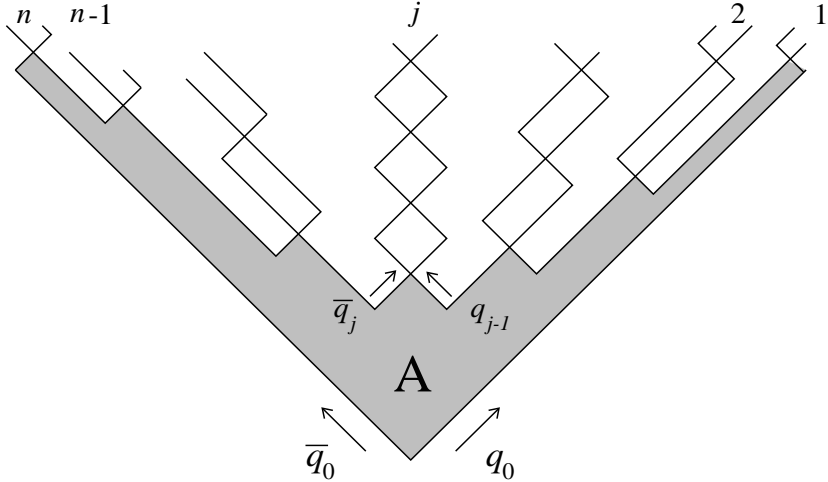,width=11cm}}
    \end{center}
  }}
\caption{\em The decay of a Lund Model string.}
\label{f:area}
\end{figure}

In the string model the confining colour field is approximated by
a massless relativistic string. The endpoints of the string are
identified with quark and antiquark properties while the gluons are assumed
to behave as transverse excitations on the string. The string can break up
into smaller pieces by the production of $q\bar{q}$-pairs (i.e. new endpoints). 
Such a pair will immediately start to separate because of the string tension, 
which in the rest frame of a string segment corresponds to a 
constant force $\kappa$;
phenomenologically $\kappa \simeq 1$ $GeV/fm$. 
Final state mesons are formed from a $q$  and a $\bar{q}$ from adjacent 
vertices, as shown
in Fig~\ref{f:area}.
	
Each breakup vertex will separate the string into two 
causally disconnected parts. From the causality, together with Lorentz
covariance and straightforward kinematics, 
it is possible to derive a unique breakup 
rule for the string by means of (semi)classical arguments \cite{ags83}.  

The unique breakup rule results in the following probability for a string to
decay into hadrons $(p_1,\dots,p_n)$.  
\eqbe
\label{e:prob}
dP(p_1,...,p_n) = \left[\prod_i (N dp_i \delta(p_i^2-m_i^2))\right] 
\delta(\sum_j p_j - P_{tot})\exp(-bA)
\eqen
where A is the area of the breakup region as indicated in Fig~\ref{f:area} 
and N and b are two parameters. 

The similarity of the result to Fermi's Golden
Rule for the probability of a quantum mechanical transition, 
i.e. the size of the final state phase space multiplied by the square of a
matrix element $|{\cal M}|^2$ expressed by the exponential area
suppression, provides a reasonable starting point to try to derive a
corresponding quantum mechanical process. There is at least two possible 
mechanisms, {\em viz.} a quantum mechanical tunneling process a la 
Schwinger and/or  the possible relationship to the Wilson loop operators in 
a gauge field theory. We will find that they provide very similar answers to
the problem.   	

Starting with the tunneling arguments, we note that while a  massless 
$q\bar{q}$-pair without transverse momentum can be produced
in a pointlike way anywhere along the string, a massive pair or a pair
with transverse momentum must classically be produced at a distance so that 
the string energy between them can be used to fulfil energy-momentum
conservation. If the
transverse momentum is conserved in the production process, i.e. the $q
\bar{q}$ with masses $\mu$ obtain $\pm \vec{k}_{\perp}$, respectively, then the
pair may classically be realised at a distance $ \delta x =
2\mu_{\perp}/\kappa$, where $\mu_{\perp}$ is the transverse mass
$\sqrt{\mu^2+\vec{k}_{\perp}^2}$. 

The probability for a quantum mechanical fluctuation of a pair, occurring with
$\mu_{\perp}$ at the (space-like) distance $\delta x$, is in a force-free region 
given by the free Feynman propagator squared:
\eqbe
\label{Feynman}
|\Delta_F(\delta x, \mu_{\perp})|^2 \sim \exp(-2 \mu_{\perp}\delta x)=
\exp\left(-\frac{4
\mu_{\perp}^2}{\kappa}\right)
\eqen
A corresponding quantum mechanical tunneling process in a constant force
field will according to WKB methods give
\eqbe
\label{e:schwinger}
\left|\exp\left(-2\int_0^{\delta x} \sqrt{\mu_{\perp}^2-(\kappa x)^2} dx \right)\right|^2
=\exp\left(-\frac{\pi \mu_{\perp}^2}{\kappa}\right) \equiv P(\mu_{\perp})
\eqen
The difference is that in the force-free case we obtain an exponential 
suppression $4\mu_{\perp}^2/\kappa$ but when the constant force pulls the pair
apart we obtain the somewhat smaller suppression $\pi\mu_{\perp}^2/\kappa$. 
Besides the mass suppression (which
phenomenologically will suppress strange quark-pairs with a factor of $\sim
0.3$ compared to ``massless'' up and down flavored pairs) we obtain 
the transverse momentum Gaussian suppression      
\eqbe
 \exp\left({-\frac{1}{2\sigma^2}k_{\perp}^2}\right) 
~~\mbox{with}~~ 2\sigma^2=\frac{\kappa}{\pi}
\label{e:pt}
\eqen
The value of $\sigma$ as used in JETSET is a bit larger than the result in
Eq~(\ref{e:pt}) but this can be understood as an effect of soft gluon 
generation along the string. The transverse momentum of a hadron produced 
in the Lund Model is then the sum of the transverse momenta of its
constituents.

We may use the elementary result in Eq (\ref{e:schwinger}) to calculate the
persistence probability of the vacuum, ${\cal P}$, 
as it is defined in \cite{gm83}.
It is the probability that the no-particle vacuum will not break up, owing to
pair-production, during the
time $T$ over a transverse region $A_{\perp}$, when a constant force $\kappa$ is
applied along the longitudinal x-direction over a region $L$:          
\eqbe
\label{totprob}
{\cal P}= \prod_{t\in(0,T), x\in(0,L), \vec{k}_{\perp}, s,f}
(1-P(\mu_{\perp}))=\exp\left(\sum_{t,x,\vec{k}_{\perp},s,f} \log(1-P)\right)
\eqen
We have then assumed that the field couples to (fermion) 
pairs with spin $s$ and
flavors $f$ and we sum over all possibilities for the production. 
As each pair needs a
longitudinal size $\delta x=2\mu_{\perp}/\kappa$ and,  
according to Heisenberg's indeterminacy
relation, will live during a timespan $2\pi/2\mu_{\perp}$  there is at most 
$\kappa LT/2\pi$ pairs possible over the space-time region $LT$. The transverse
momentum summation can be done by Gaussian integrals from an expansion of 
$\log(1-P)$
and the introduction of the well-known  number of waves available 
in a transverse
region $A_{\perp}$: $(A_{\perp}/(2\pi)^2) d^2k_{\perp}$. In
this way we obtain for the persistence probability
\eqbe
\label{finalprob}
{\cal P} = \exp(-\kappa^2 L T A_{\perp} \Pi) ~~~ \mbox{with} ~~~ \Pi=\frac{n_f
n_s}{4 \pi^3} \sum_{n=1}^{\infty} \frac{1}{n^2} \exp\left(-\frac{n \pi
\mu^2}{\kappa}\right)
\eqen
where $n_f,n_s$ is the number of flavor and spin states. 

There are two remarks to this result. Firstly, although the method to treat 
the integration over time and longitudinal space, by close-packing 
reasonably sized boxes, may not seem convincing the final
formula \cite{gm83} coincides with the one obtained by Schwinger 
\cite{js51}, for the
case of a constant electric field ${\cal E}$. Then 
$\kappa$ is identified with 
the force of the charges in the external field, i.e. $\kappa \rightarrow 
e {\cal E}$. 

Secondly, the result is in evident
agreement with the formula for the decay of the Lund string in Eq
(\ref{e:prob}) if we identify $ L T$ with the (coordinate space)
area size $A$. In this way we also obtain the result that the parameter $b$ is
\eqbe
\label{binterprete} 
b=\kappa^2 A_{\perp} \Pi
\eqen 
i.e. it corresponds to the transverse size of the 
(constant)
force field, which we have modelled by the string. The quantity $\Pi$ is 
$1/(12 \pi)$ for two massless spin $1/2$-flavors. 

The second quantum mechanical approach is to note that a final state hadron
stems from a $q$ from one vertex $j$ and a $\bar{q}$ from the adjoining vertex
$j+1$. In
order to keep to gauge invariance it is then necessary that the production     
matrix element contains at least a gauge connector between the vertices: $\exp
(i \int_j^{j+1} g {\cal A}^{\mu} dx_{\mu})$, where $g$ is the charge and ${\cal
A}^\mu$ the gauge field. Consequently the total production matrix element must
contain a Wilson loop operator:
\eqbe
\label{Wilsonloop}
{\cal M} = \exp(i \oint g {\cal A}^{\mu} dx_{\mu})  
\eqen
with the integration around the region $A$ (note that the field is singular
along the border line and we are therefore not allowed to distort the
integration contour inwards). The operator in Eq (\ref{Wilsonloop}) was
predicted (and inside lattice gauge calculations also found) to behave as 
\eqbe
\label{Wilsonarea}
{\cal M}= \exp(i \xi A)
\eqen
with the real part of $\xi$, $Re(\xi)=\kappa$. In the present situation where
the force field region decays we expect an imaginary part,
corresponding to the pair production rate according to the well-known
Kramers-Kronig \cite{kramerskronig} relationship for the dielectricity in
matter, in this case the QCD vacuum.

The two interpretations of the area law, i.e. the Schwinger tunneling in Eq
(\ref{finalprob}) and the Wilson loop operator result in Eq
(\ref{Wilsonarea}) can be related if we note that according to Gauss' Law the
integral over the extension of the force field should correspond to the charge.
For a thin string we should then obtain for the area falloff rate 
$b \propto \kappa^2 A_{\perp} \propto
\kappa \alpha$. Although Gauss' law is more complicated for 
a non-abelian field with
triplet and octet color-charges and similarly octet fields it is possible to
make a case for an identification of the parameter $b$ as
\eqbe
\label{binterprete2}
b = \frac{\kappa n _f\bar{\alpha}}{12}
\eqen
which is what we should expect from the expected imaginary part of the
dielectricity in Eq (\ref{Wilsonarea}). $\bar{\alpha}=3g^2/(4\pi)$ is then the effective QCD
coupling, including the color factors.    
The result is also phenomenologically supported if we consider a partonic
cascade down to a certain transverse momentum cutoff $k_{\perp c}$ and then use
the Lund model hadronisation formulas to obtain the observed properties of the
final state. In that way we 
may determine the parameters in the model as functions of the partonic cascade
cutoff. A remarkably good fit to the $b$-parameter is given by
$C/\log(k_{\perp c}^2/\Lambda^2)$ with $C$ given by Eq (\ref{binterprete2}) and
$\Lambda \simeq .5$ $GeV$ \cite{cs91}, according to the QCD coupling.

Independently of the precise identification of $b$, we obtain a possible matrix
element from Eq (\ref{Wilsonarea})
\eqbe
\label{matrixelement}
{\cal M}= \exp(i\kappa -b/2)A
\eqen
which not only will provide us with the Lund decay probability in Eq
(\ref{e:prob}), but also can be used in accordance with \cite{ah86} to
provide a model for the Hanbury-Brown-Twiss effect (often called the
Bose-Einstein effect) for the correlations among identical bosons.

\section{A Model for Bose-Einstein Correlations} \vspace{-2ex}
We will from now on work in energy-momentum space in agreement with the usual
treatment of the Lund model formulas. Further we will make use of a lightcone
metric with $p_{\pm} = e\pm p_{\ell}$ where $\ell$ denotes the longitudinal direction along the string. The two metrics differ by a factor of two, i.e.  
$2 de dp_{\ell} = dp_+ dp_-$.
Note in particular that compared to the considerations in 
the earlier section this means that the area $A \rightarrow 2\kappa^2 A$ and the parameter $b \rightarrow b/(2\kappa^2)$.

We now consider a final state containing (among possibly a lot
of other stuff) $n$ identical bosons.
There are $n!$ ways to produce such a state, each corresponding to
a different permutation of the particles.
According to quantum mechanics the transition matrix element is to be 
symmetrised with respect to exchange of identical bosons. This leads to the
following general expression for the production amplitude
\eqbe
{\cal M} = \sum_{{\cal P}}{\cal M}_{\cal{P}}
\eqen
where the sum goes over all possible permutations ${\cal P}$ of the 
identical bosons. The cross section will then contain the 
square of the symmetrised amplitude ${\cal M}$  
\eqbe  
|{\cal M}|^2 = \sum_{{\cal P}}(|{\cal M}_{\cal{P}}|^2 (1 + 
\sum_{{\cal P}^{\prime}\neq {\cal P}} \frac{2 \mbox{Re}({\cal M}_{{\cal P}} 
{\cal M}_{{\cal P}^{\prime}}^*)}{|{\cal M}_{{\cal P}}|^2 + 
|{\cal M}_{{\cal P}^{\prime}}|^2}))
\label{e:symM2} 
\eqen 
The phenomenological models used to describe the hadronisation process are 
formulated in a probalistic language (and not in an amplitude based language).
This implies that interference between different ways to produce identical 
bosons is not included. In this case the probability for producing the 
state is 
\eqbe
|{\cal M}|^2 = \sum_{{\cal P}}|{\cal M}_{\cal{P}}|^2 
\label{e:M2}
\eqen
instead of the probability in Eq~(\ref{e:symM2}). 
Comparing Eq~(\ref{e:symM2}) and Eq~(\ref{e:M2}) it is seen that a particular
production configuration leading to the 
final state ${\cal P}$ can be produced according to a probalistic scheme and 
that the quantum mechanical interference from production of identical bosons 
can be incorporated by weighting the produced event with
\eqbe
w =1+\sum_{{\cal P}^{\prime}\neq {\cal P}} \frac{2 \mbox{Re}({\cal M}_{{\cal P}} 
{\cal M}_{{\cal P}^{\prime}}^*)}{|{\cal M}_{{\cal P}}|^2 + 
|{\cal M}_{{\cal P}^{\prime}}|^2}
\eqen 
The outer sum in Eq~(\ref{e:symM2}) is as usual taken care of by generating 
many events.
\begin{figure}[t]
  \hbox{\vbox{
    \begin{center}
    \mbox{\psfig{figure=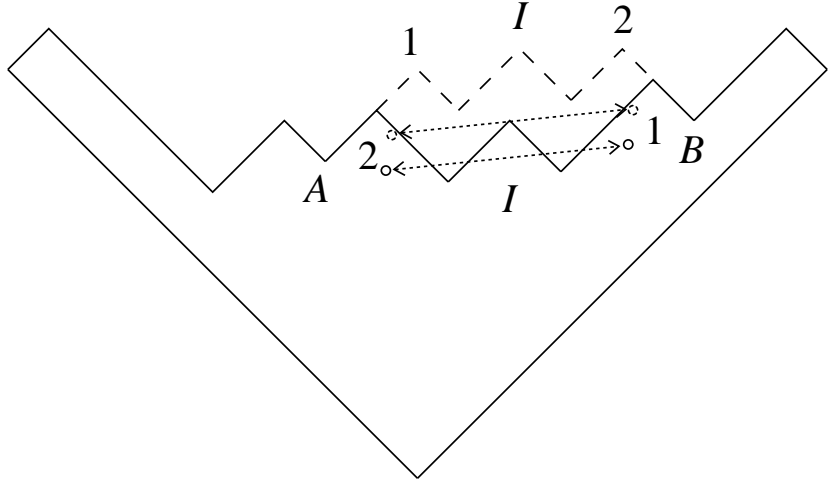,width=11.0cm}}
    \end{center}
  }}
\caption{\em The two possible ways to produce the entire state when two of the bosons are identical.}
\label{f:areadiff}
\end{figure}

In order to see the main feature of symmetrising the hadron production
amplitude in the Lund Model we consider Fig~\ref{f:areadiff}, in which two
of the produced hadrons, denoted $(1,2)$, are assumed to be identical bosons
and the state in between them is denoted $I$. 
We note that there are two different ways to produce the entire state 
corresponding to the production configurations $(\ldots,1,I,2,\ldots)$ 
and $(\ldots,2,I,1,\ldots)$, i.e. to exchanging the two identical bosons.
The two production configurations are shown in the figure and the main 
observation is that they in general correspond to different areas! 

The area difference, $\Delta A$, depends only on the energy momentum
vectors $ p_1, p_2$ and $ p_I $, but can in a 
dimensionless and intuitively useful way be written
\eqbe
\frac{\Delta A}{2\kappa} = \delta p \delta x 
\label{e:deltaA}
\eqen
where $\delta p = p_2-p_1$ and $\delta x = (\delta t; 0, 0, \delta z)$ 
is a reasonable estimate of the
space-time difference, along the surface area, between the production points
of the two identical bosons. We note that
the space-time difference $\delta x$ is always space-like. In 
Fig~\ref{f:areadiff} $\delta x$, for the two production configurations, 
is indicated by arrows, together with 
open circles showing the corresponding production points.

We go on to consider the effects of transverse momentum generation 
in the $q\bar{q}$-vertices. First we note that the total transverse momenta 
of the substate $1,I,2$ in 
Fig~\ref{f:areadiff} stem from the $q$ and $\bar{q}$  
generated at the two surrounding vertices, $A$ and $B$. This 
is, owing to momentum conservation, fixed by the properties of 
the hadrons generated outside of 
the substate. 
Using this we find 
that there is a unique way to change the transverse momenta in the 
vertices surrounding the intermediate state $I$ such that
every hadron has the same transverse momenta in both production 
configurations. 

Suppose as an example that we have generated $\pm {\bf k}_{\perp A}$ 
in the vertex $A$ and $\pm {\bf k}_{\perp B}$ in the vertex $B$ 
(i.e. so that $-{\bf k}_{\perp A}$ and ${\bf k}_{\perp B}$ defines the 
substate). Then to conserve 
the transverse momenta of the observed hadrons when changing
production configuration from $(1,I,2)$ to $(2,I,1)$ it is necessary to 
change the generation of transverse momenta in the two vertices surrounding
$I$ as follows (in an easily understood notation):
\eabe
\pm {\bf k}_{\perp I} \rightarrow \pm ({\bf k}_{\perp A}+{\bf k}_{\perp B}-
 {\bf k}_{\perp I}^{\prime}) \\
\pm {\bf k}_{\perp I}^{\prime} \rightarrow \pm ({\bf k}_{\perp A}+
 {\bf k}_{\perp B}-{\bf k}_{\perp I}) \nonumber
\eaen
This means that exchanging two bosons with different transverse momenta
will result in a change in the amplitude as given by Eq~(\ref{e:pt}) for
some of the vertices. 

From the amplitudes in Eq~(\ref{matrixelement}) and 
Eq~(\ref{e:pt}) we get that the weight in the Lund Model can be written	
\eqbe
 w = 1+\sum_{{\cal P}^{\prime}\neq {\cal P}}\frac {\textstyle \cos \frac{\textstyle \Delta A}{\textstyle 2\kappa}}
 {\textstyle \cosh \left( \frac{\textstyle b\Delta A}{\textstyle 2}+
 \frac{\textstyle \Delta(\sum p^{2}_{\perp q})}{\textstyle 2\sigma^{2}_{p_\perp}}\right)}
\label{e:weight} 
\eqen
where $ \Delta $ denotes the difference with respect to the configurations 
${\cal P} $ and ${\cal P^{\prime}}$ and the sum of $p^{2}_{\perp q}$
is over all vertices. We have introduced $\sigma_{p_\perp}$ as the 
width of the transverse momenta for the generated hadrons, (i.e. 
$\sigma^{2}_{p_\perp}= 2\sigma^2$).

Using Eq~(\ref{e:deltaA}) for a single pair exchange one sees that the area 
difference is, for small $\delta p$, governed by the distance between 
the production points and that $\Delta A$ increases quickly with this distance.
We also note that $\Delta A$ vanishes with the four-momentum difference
and that the contribution to the weight from a given configuration, 
${\cal P}^{\prime}$, vanishes fast with increasing area difference $\Delta A$.
From these considerations it is obvious that only exchanges of pairs with 
a small $\delta p$ and a small $\delta x$ will give a contribution to the 
weight. In this way it is possible to relate to the
ordinary way to interpret the HBT effect, cf. Eq (\ref{calRform}). 

It is straightforward to generalise Eq~(\ref{e:deltaA}) to higher order 
correlations. One notes in particular that the area difference does not 
vanish if more than two identical bosons are permuted 
and only two of the bosons have identical four-momenta. 

Models for BE correlations has been suggested, e.g. \cite{st97}, with similar 
weight functions, but it is important to note that the weight in our model
has a scale both for the argument to the cos-function as well as for the 
function which works as a cut-off for large $\delta p$ and $\delta x$ 
(in our case a cosh-function). Further the two scales in our model 
are different and well-defined, at least phenomenologically. 
We will come  back to the influence of the two scales 
in Section~\ref{s:results}. 
\label{s:model}

\section{MC Implementation} \vspace{-2ex}
To calculate the weight for a general event, with multiplicity n, 
one has to go through $ n_{1}!n_{2}!...n_{N}!-1 $  
possible production configurations, where $ n_{i} $ 
is the number of particles of type i and there are 
N different kinds of bosons. For a general $ e^{+}e^{-}$-event at 
90 GeV this is not possible from a computational point of view.

We know however that the vast majority of configurations will give large
area differences and they will therefore not contribute to the weight.
One of our aims with this work has been to find a way to approximate the sum 
in Eq~(\ref{e:weight}) with {\em a sum over configurations with significant 
contributions to the weight.} From basic group theory we know that every 
group can be partitioned into its classes. Let $11111\ldots1$
denote the class containing only the identity element, where all particles 
are unchanged, $m_1m_2111\ldots 1$ denote 
the class of group-elements where $m_1$ particles are cyclically permuted, 
$m_2$ other particles are cyclically permuted and
the rest are unchanged, and so on. We can define the order, $k$, of a class 
as $k=\sum_i(m_i-1)$. 
The useful feature of this ordering of classes is that for 
all group-elements contained in order $k$ the minimum of the summed size 
(in positions) of the cyclically permuted clusters , $\Delta r$, is $k$, 
i.e. the minimum length over which particles are moved increases with 
the order. 
From the discussion at the end of Section~\ref{s:model} it is then obvious 
that the contribution to the weight from a configuration will decrease with 
its order. All classes up to order 4 are shown in Table~\ref{t:classes}. 
 
\begin{table}[ht]
\cebe
\begin{tabular}{|c|c|c|c|c|c|}
\hline
Order & 0 & 1 & 2 & 3 & 4 \\ \hline 
Classes &  11111\ldots 1 & 21111\ldots 1 & 22111\ldots 1 & 22211\ldots 1 & 22221\ldots 1 \\ 
 & & &  31111\ldots 1 & 32111\ldots 1 & 32211\ldots 1 \\ 
 & & & & 41111\ldots 1 & 33111\ldots 1 \\ 
 & & & & & 51111\ldots 1 \\ 
 & & & & & 42111\ldots 1 \\ 
\hline
$\Delta r$ & 0 & $\geq$ 1 & $\geq$ 2 & $\geq$ 3 & $\geq$ 4 \\
\hline
\end{tabular}
\ceen
\label{t:classes}
\caption{\em The classes of the permutation group order by order up to the fourth order. $\Delta r$ is the minimum length over which particles are permuted.}
\end{table}
We have found that for essentially all events the weight does not change when 
including the fifth order. But we have also found that lots of 
lower-order configurations give no contributions to the weight.
This is not acceptable when taking computing time into account and we have
therefore abandoned using a cut in order.

In this work we have instead approximated the sum in Eq~(\ref{e:weight})
with a sum over configurations of all orders with significant contributions 
to the weight. 
This has been done by introducing {\em exchange-links} between particles.
We have only taken into account interference with configurations where 
all particles are produced in positions from which there is a link to a
particle's original production position. Defining an exchange matrix, 
$ {\cal L} $, as follows

\[ {\cal L}_{ij} = \left\{ \begin{array}{ll}
			    1 & \mbox{if there is a link between 
				      particles i and j.} \\
		            0 & \mbox{otherwise}
			   \end{array}
                   \right. \]
one gets a simple representation of the configurations to be 
considered. If all elements in ${\cal L}$ are 1, it corresponds to considering
all n! permutations, while only the original configuration is considered 
if ${\cal L}$ is the identity matrix. 

$ \Delta A $ for a pair exchange increases, as previously discussed, with 
the four-vector difference $ \delta p $ and with the size of the state 
inbetween. 
Since we know from Eq~(\ref{e:weight}) that the contribution to the weight 
for a given configuration vanishes fast with increasing area difference 
$ \Delta A $, it is useful to introduce the concept of linksize, 
defined below as the invariant four-momentum difference together with 
the invariant mass of the particles 
produced in between the pair (in rank). 
By only accepting links 
between particles if the size of the link between them is smaller than
some cut-off linksize, $\delta_c$, we get a prescription for the exchange 
matrix of an event.
In this way, by specifying the allowed two-particle exchanges, we get, 
to all orders, which configurations to take into account.
We have found that for a given $\delta_c$ one includes all configurations
that provide a contribution larger than some $\epsilon$ to the weight. 
Taken together this means that we get all the important contributions to 
the weight if we chose $\delta_c$ so large that
the neglected terms smaller than $\epsilon$ give a negligable change
for every weight. 

We have used a cut-off linksize such that there is a link between identical bosons if one of the following conditions is fulfilled. 
\itbe
\item $Q^2= -(p_i-p_j)^2 < Q_{max}^2 \simeq 1~GeV^2 $.
\item the invariant mass of the particles produced inbetween 
(along the string) the pair is less than $m_{max}^2 \simeq (20~GeV)^2 $.
\iten
Including links larger than this give no contribution to the weight 
for essentially all events. 
There are a few special events for which the weights have not converged 
with this $\delta_c$. They are very rare and have in common that they 
have a cluster of particles such that exchanging any pair in the cluster will
give a large area-difference, but there are cyclic permutations which give
a small area-difference. Increasing $\delta_c$ to include these configurations
give no noticeable effect in any observable known to us (except the computing
time in the simulation!). 

\subsection*{Including decays}
A large fraction of all final-state bosons stem from decays of short-lived 
resonances with lifetimes comparable to the time scale in string decay. 
Therefore they may contribute to the Bose-Einstein effect. To include their 
decay amplitudes and phase space factors and symmetrise the total amplitude
is very difficult and it is furthermore not known how to do that in a 
model-consistent way. We have included resonance decays in the following 
simple way
\begin{quote}
 Particles with width larger than $\Gamma_{min}$ are assumed to decay 
before Bose-Einstein symmetrisation sets in and the matrix elements
are evaluated with their decay products regarded as being 
produced directly, ordered in rank. We have used $\Gamma_{min} = 0.02$ GeV. 
\end{quote}
The signal in the two-particle correlation function goes down very much if 
we neglect all the pions from resonance decays when symmetrising the 
amplitudes. But our signals are fairly independent 
of $\Gamma_{min}$ as long as it is small enough for the $\rho$'s
to decay before  the symmetrisation.

An elaborate discussion on the treatment of resonances in connection with 
BE correlations can be found in \cite{mgb86}.

\section{Results} \vspace{-2ex}
\label{s:results}
In our simulations we have used the Lund string model \cite{ba83} implemented in the JETSET MC \cite{ts94} to hadronize $ \epq\epqa $-pairs (i.e.
no gluons are considered). The MC implementation of our model is available
from the authors.

\subsection{The Weight Distribution and Two-particle Correlations}

\begin{figure}[t]
  \hbox{\vbox{
    \begin{center}
    \mbox{\psfig{figure=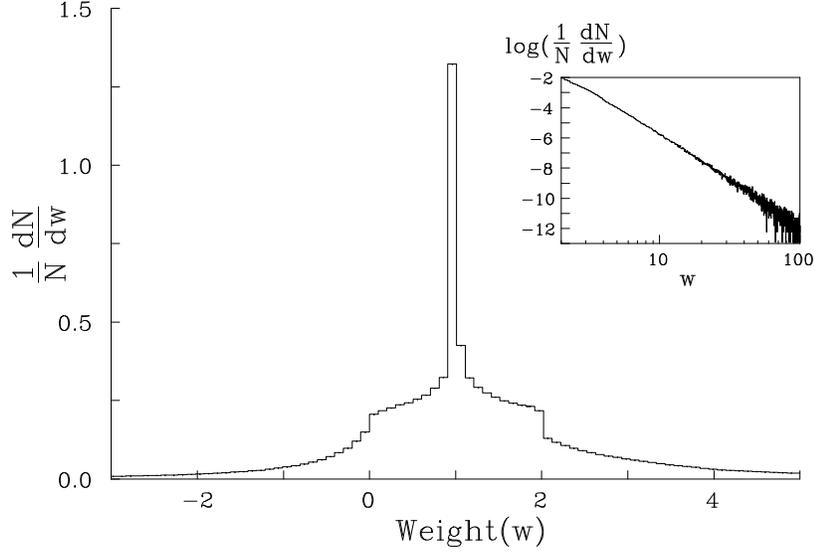,width=12cm}}
    \end{center}
  }}
\caption{\em The distribution of Bose-Einstein weights for two-jet states in 
	 JETSET. The tail of positive weights is shown in the insert.}
\label{f:weight}
\end{figure}

\begin{figure}[t]
  \hbox{\vbox{
    \begin{center}
    \mbox{\psfig{figure=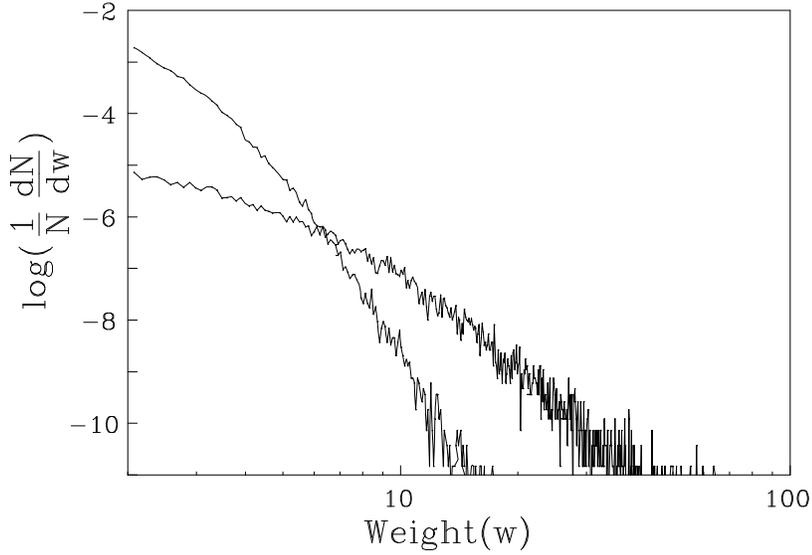,width=12cm}}
    \end{center}
  }}
\caption{\em The distribution of Bose-Einstein weights for two-jet events 
	 subdivided into sets with different number of links, $n_l$.
 	 Two samples for $ 3\leq n_l \leq 5 $ and $ 10 \leq n_l \leq 12 $ are
	plotted.}
\label{f:weight_links}
\end{figure}

The majority of the weights are close to and centered around unity, as seen
in Fig~\ref{f:weight}. There is however a tail of weights far away from
unity in both directions. The tail of positive weights is shown as an insert 
and the distribution looks like a power. However if we subdivide the events 
into sets with similar number of links and study the weight distributions for
these sets separately, we find that the weight distribution for each set is
basically Gaussian. The width of these Gaussians increase with the number 
of links in the corresponding set, as shown in Fig~\ref{f:weight_links}.
The power like behaviour of the weight distribution is therefore merely a
consequence of summing over events with different number of links. It should 
be emphasized that the negative weights only are a technical problem.
Summing over many events results in positive probabilites for all physical
observables, which is obvious from Eq~(\ref{e:symM2}).   


We have taken the ratio of the two-particle probability density of pions,
$\rho_2$, with and without BE weights applied as the two-particle correlation 
function, $R_2$, i.e.
\eqbe
R_2(p_1,p_2) = \frac{\rho_{2w}(p_1,p_2)}{\rho_2(p_1,p_2)}
\eqen
where the $w$ denotes weighted distributions.

\begin{figure}[h,t,b]
  \hbox{\vbox{
    \begin{center}
    \mbox{\psfig{figure=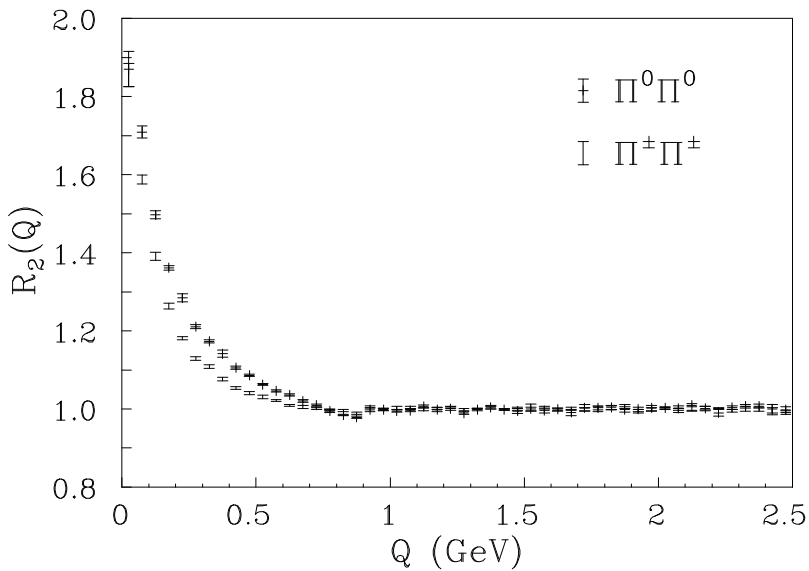,width=12cm}}
    \end{center}
  }}
\caption{\em The ratio $R(Q)$ of the number of pairs with invariant relative 
	 four-momentum $Q$ with and without Bose-Einstein weights applied. The 
         sample consists of the initial particles used in the symmetrisation.}
\label{f:R_Q}
\end{figure}

As discussed in connection with Eq~(\ref{e:deltaA}) 
the correlation length in Q depends inversely on the (space-like) 
distance between the production points of the identical bosons and
the Bose-Einstein correlation length, that is dynamically 
implemented, in this model can most easily be described as the flavour 
compensation length, i.e.
the region over which a particular flavour is neutralised. 
Identically charged particles cannot be produced as neighbours along the 
string in the Lund model while neutral particles can.
This implies that identically charged 
pions which always must have a non-vanishing state inbetween will have a 
more narrow correlation distribution in Q compared to neutral pions. 
This has been found as can be seen in Fig~\ref{f:R_Q} where the correlation
distributions for pairs of particles used in the symmetrisation are shown.
The correlation functions have been normalised to unity in the 
region $1.0\leq Q\leq 2.5$.
The correlation distribution for charged pions can be approximated
by the LUBOEI algorithm \cite{ls95} with radii $\simeq 1$ fm and 
$\lambda \simeq 0.8$ as
input parameters (Note that the input parameters are not exactly reproduced
 in the resulting correlation function). 
This is in reasonable agreement with the LEP experiments,
which measure sizes of the order of $0.5-1~fm$, \cite{aleph92,delphi94,opal91},
even though they tend towards values smaller than $1~fm$. The correlation 
function depends on the daughters of resonances and especially the decay 
products of $\eta'$ play a large role. The production rate of $\eta'$
used in JETSET was questioned in \cite{mgb86}, in connection with 
BE correlations. We have used reduced production rates for $\eta$ and 
$\eta'$ by setting the extra supression factors in JETSET to 0.7 and 0.2 
respectively, in accordance with the DELPHI tuning \cite{DELPHItuning95}.
For a more elaborate quantative comparison with data our BE Monte Carlo
has to be tuned further and the resulting events has to be subjected to 
the same corrections as in the experimental analysis.

The general findings for the parameter dependence of the weight function in Eq~(\ref{e:weight}) is that due to the smallness of $b$ as compared to $1/\kappa$,
it is for most of the terms in the sum the decrease of the cos-function with
increasing $\Delta A$ that governs the behaviour. For larger $\Delta A$ it is
the transverse momentum contributions to the cosh-function which takes over
to damp the contribution to the weight. Note that the argument of the 
cos-function contains $\kappa$ as the basic scale and that the transverse 
momentum contributions also are governed essentially by $\kappa$ (see Eq~(\ref{e:pt})). Going over to correlation functions we find as expected from the
conclusions for the weight function that the correlation function is 
not affected when the $b$-parameter is changed $\pm$20\%. The slope of the
correlation function for small $Q$ values and therefore the correlation length
is very sensitive to $\kappa$ and it is also sensitive to the width of the 
transverse momenta. The transverse momentum generation acts as noise in the 
model so that all weights approach unity and consequently all correlations
vanish with increasing $\sigma_{p_\perp}$.
It is however this noise which makes the weight calculations tractable.
Consequently, the main parameter  is the string tension, $\kappa$, 
in this model for Bose-Einstein correlation weights as well as for 
the correlation length.

\begin{figure}[t]
  \hbox{\vbox{
    \begin{center}
    \mbox{\psfig{figure=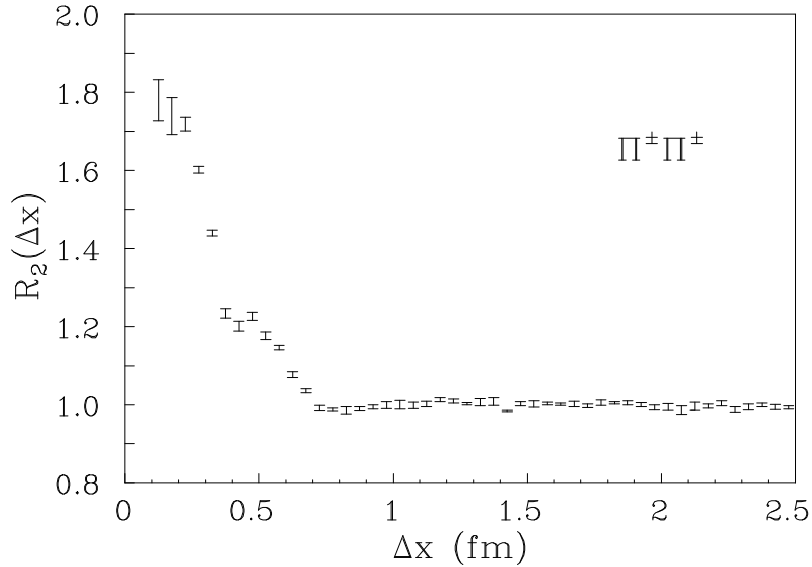,width=12cm}}
    \end{center}
  }}
\caption {\em The ratio $R_2(\Delta x$) of the number of pairs with invariant relative distance  $\Delta x$ with and without Bose-Einstein weights applied. The sample consists of the initial particles used in the symmetrisation.}
\label{f:R_deltax}
\end{figure}

Since the BE weights are depending on the space-like distance between the 
production points we have studied the two-particle correlation function as 
a function of the invariant space-time distance $\Delta x = 
\sqrt{-\delta x^2}$ where $\delta x = (\delta t; 0, 0, \delta z)$
as defined in Eq~(\ref{e:deltaA}). 
In Fig~\ref{f:R_deltax} $R_2(\Delta x)$, which has
been normalised to unity in the region $1.0 \leq Q \leq 2.5$, is plotted. 
The figure illustrates that the effect of the Bose-Einstein symmetrisation
, i.e. to pack identical bosons closer together in phase-space, is manifest 
up to production point separations of about 0.7 $fm$. It should however be 
noted that many configurations where pairs are exchanged over significantly 
larger distances give significant contributions to the weight.

We have also found that the higher order contributions to the sum in 
Eq~(\ref{e:weight}) is of importance for the two-particle correlations.
That is using more than two-particle exchanges when calculating the weights 
does not only affect the weight distribution but also the
two-particle correlation function, $R_2(Q)$.

In heavy-ion collision experiments one has found that the extracted 
correlation length has an approximate $ 1/\sqrt{m_t} $ dependence \cite{naXX}. 
This is in agreement with hydrodynamical models describing the 
source evolution in heavy-ion collisions.
Recently a similar $m_t$ dependence has been found for $Z^0$ hadronic decays
in $e^+e^-$ annihilation at LEP \cite{ls96}.
In the Lund Model the average space-like distance between pairs of identical
pions increases with $m_t$ and one would therefore not expect a correlation 
length which falls off with $m_t$. For initially produced particles we get a correlation length which is essentially independent of $m_t$. However when 
analysing all final particles we find for increasing $m_t$ that the correlation length falls off and that the $\lambda$ parameter increases, as in \cite{ls96}.
From this we conclude that the observed $m_t$ dependence of the correlation length in data is, in our model, compatible with the vanishing of contributions from
decay products with increasing $m_t$.

\subsection{Residual Bose-Einstein correlations}
Bose-Einstein correlations acting between identical bosons may have significant
indirect effects on the phase space for pairs of non-identical bosons. We have 
studied mass distributions of $\pi^+\pi^-$ systems to see how our model 
affects systems of unlike charged pions. Many analyses use $\pi^+\pi^-$ 
distributions to quantify the Bose-Einstein correlations, using the unlike-charged distributions as reference samples with which to compare the like-charged 
pion distributions. We have found that the assumption that the two-particle 
phase space densities for $\pi^+\pi^-$ systems are relatively unaffected by
Bose-Einstein symmetrisation is fairly good. Taking the ratio of the $\pi^+\pi^-$ mass distributions  with and without Bose-Einstein symmetrisation applied 
gives that the mass distribution is not altered much by the symmetrisation,
and that the effect is smaller than 5\% in the entire mass range.
\begin{figure}[h,t]
  \hbox{\vbox{
    \begin{center}
    \mbox{\psfig{figure=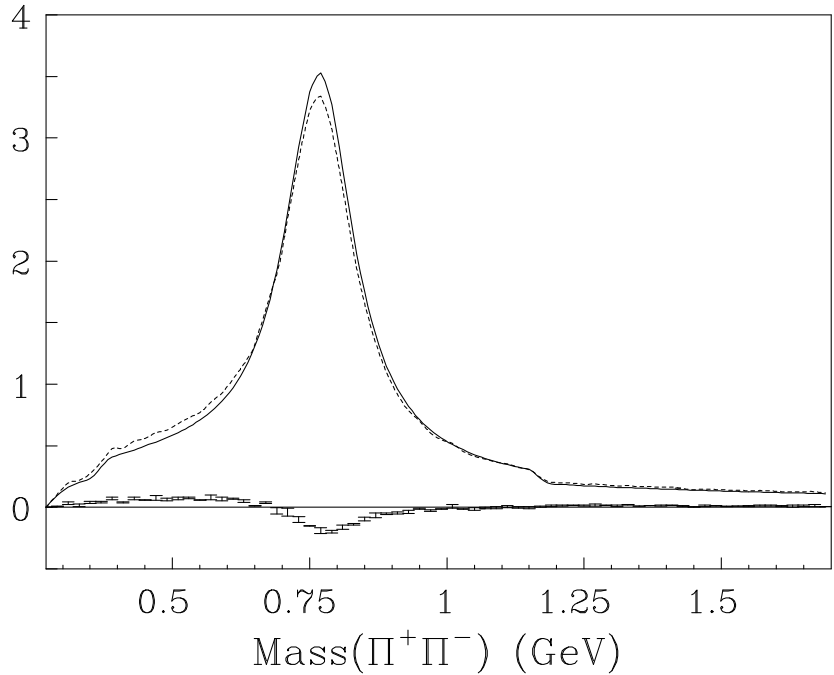,width=12cm}}
    \end{center}
  }}
\caption {\em $\rho^0$ meson mass shift induced by the Bose-Einstein 
          correlations in our model. The solid curve shows the Breit-Wigner 
          as generated by JETSET while the dashed curve is obtained after
	  applying BE weights to the events. The curve with error bars is the
	  difference of the two (dashed-solid). The areas under the two mass 
          distributions are normalised to unity in the shown mass range.}
\label{f:rhomass} 
\end{figure}

It has however been observed experimentally that 
the Breit-Wigner shape for oppositely charged pions from the decay of the $\rho$
resonance \cite{opaldelphirho,delphirho94,gl93} is distorted. 
We have therefore analysed
$\pi^+\pi^-$ distributions when the pair comes from the decay of a $\rho^0$.
These $\pi^+\pi^-$ mass distributions, with and without BE weights 
applied, are shown together with the difference of the two in 
Fig~\ref{f:rhomass}. From the difference it is clearly seen that the weighting 
depletes the region around the $\rho$ mass and shifts the masses towards lower 
values as well as it slightly increases the width of the distribution. 
The figure clearly shows the potential of our model to affect the mass
spectrum of the $\rho^0$.

\subsection{Three-particle correlations}
The existence of higher order dynamical correlations, which are not a 
consequence of 
two-particle correlations, is of importance for the understanding 
of BE correlations. There are very few
experimental studies of genuine three-particle correlations, mainly because
of the problem of subtracting the consequenses of two-particle correlations 
and the need for high statistics of large multiplicity events. 
Genuine short-range three-particle correlations have been observed in 
$e^+e^-$ annihilations by the DELPHI experiment. They conclude that
they can be explained as a higher order Bose-Einstein effect \cite{3corr95}.

To reduce problems with pseudo-correlations due to the summation of events 
with different multiplicities we have used three-particle densities normalised
to unity separately for every multiplicity in the following way 
\eqbe
\tilde{\rho}_3^{(a,b,c)}(p_1,p_2,p_3) = \sum_{n \geq 8}P(n_a,n_b,n_c)\tilde{\rho}^{(n_a,n_b,n_c)}_3(p_1,p_2,p_3)
\label{e:rhotilde3}
\eqen
\eqbe
\tilde{\rho}^{(n_a,n_b,n_c)}_3(p_1,p_2,p_3) = \frac{1}{n_a(n_b-\delta_{ab})(n_c-\delta_{ac}-\delta_{bc})}\frac{1}{\sigma_{(n_a,n_b,n_c)}}\frac{d^3 \sigma_{(n_a,n_b,n_c)}}{dp_1dp_2dp_3}
\label{e:rhotilde3n}
\eqen
where $n$ is the charged multiplicity, $\sigma_{n_a,n_b,n_c}$ is the semi-inclusive cross section for events with $n_i$ particles of species $i$, and
\eqbe
P(n_a,n_b,n_c)= \frac{\sigma_{(n_a,n_b,n_c)}}{\sum_{n_a,n_b,n_c}\sigma_{(n_a,n_b,n_c)}}
\eqen
We have aimed to study the genuine normalised three particle correlation function, 
$\tilde{R}_3$, defined as
\eabe
\tilde{R}_3 & = & [\tilde{\rho}_{3}(p_1,p_2,p_3)
                -\tilde{\rho}_2(p_1,p_2)\tilde{\rho}_1(p_3)
                -\tilde{\rho}_2(p_1,p_3)\tilde{\rho}_1(p_2)
                -\tilde{\rho}_2(p_2,p_3)\tilde{\rho}_1(p_1) \nonumber \\ 
         & & + 2\tilde{\rho}_1(p_1)\tilde{\rho}_1(p_2)\tilde{\rho}_1(p_3)]
             /(\tilde{\rho}_1(p_1)\tilde{\rho}_1(p_2)\tilde{\rho}_1(p_3))+1
\label{e:Rtilde3}
\eaen
where we have used an abbreviated notation for the $\tilde{\rho}_3$ from
Eq~(\ref{e:rhotilde3}), and $\tilde{\rho}_1$ and $\tilde{\rho}_2$ are the 
corresponding one- and two-particle densities, normalised in accordance 
with Eq~(\ref{e:rhotilde3}) and Eq~(\ref{e:rhotilde3n}). 
$\tilde{R}_3$ is equal to one if all three-particle correlations are consequenses of two-particle correlations.

In order to calculate the $\tilde{\rho}_2\tilde{\rho}_1$ and $\tilde{\rho}_1\tilde{\rho}_1\tilde{\rho}_1$ terms in Eq~(\ref{e:Rtilde3}) the common experimental
procedure is to mix tracks from different events. Using a mixing procedure
in our model means weighting triplets of particles with products of event 
weights. This results in large statistical fluctuations and to get them 
under control, with our event weights, requires generation of very many events. 
We have therefore taken another approach, in order to minimise the 
computing time. We have used combinations of 
charged pions in the following way to approximate Eq~(\ref{e:Rtilde3}) 

\eqbe
\tilde{R}_3 \equiv \frac{\tilde{\rho}_{3w}^{(\pm,\pm,\pm)} - 3(\tilde{\rho}^{(\pm,\pm,\mp)}_{3w} - \tilde{\rho}_3^{(\pm,\pm,\mp)})}{\tilde{\rho}_{3}^{(\pm,\pm,\pm)}}
\label{e:Rtilde3appr}
\eqen
where $w$, as previously, denotes weighted distributions. There are a couple of
things to note in connection with Eq~(\ref{e:Rtilde3appr}). If there are 
genuine positive three-particle correlations for $(++-)$ 
and $(--+)$ combinations, as observed by the DELPHI collaboration \cite{3corr95}
 they will if they come from BE symmetrisation contribute to the $\tilde{R}_3$ 
in Eq~(\ref{e:Rtilde3appr}), but they will reduce the signal. Secondly, we note 
that there is a possible bias from two-particle correlations from $(+-)$ 
combinations but that it is small as discussed previously. We also note that
using the normalisation in Eq~(\ref{e:rhotilde3n}) reduces problems with
contributions from like- and unlike-charge combinations having different 
multiplicity dependence. It should also be observed that the $\tilde{R}_3$
in Eq~(\ref{e:Rtilde3appr}) can be studied experimentally since 
getting the $\tilde{\rho}_{3w}$'s of course is achieved by analysing single
events and the $\tilde{\rho}_{3}$ samples can be made by mixing events.  

We have analysed the three-particle correlations as a function of the 
kinematical variable
\eqbe
Q=\sqrt{q_{12}^2+q_{13}^2+q_{23}^2} ~~~~\mbox{with} ~~~~q_{ij}^2=-(p_i-p_j)^2
\eqen

\begin{figure}[t]
  \hbox{\vbox{
    \begin{center}
    \mbox{\psfig{figure=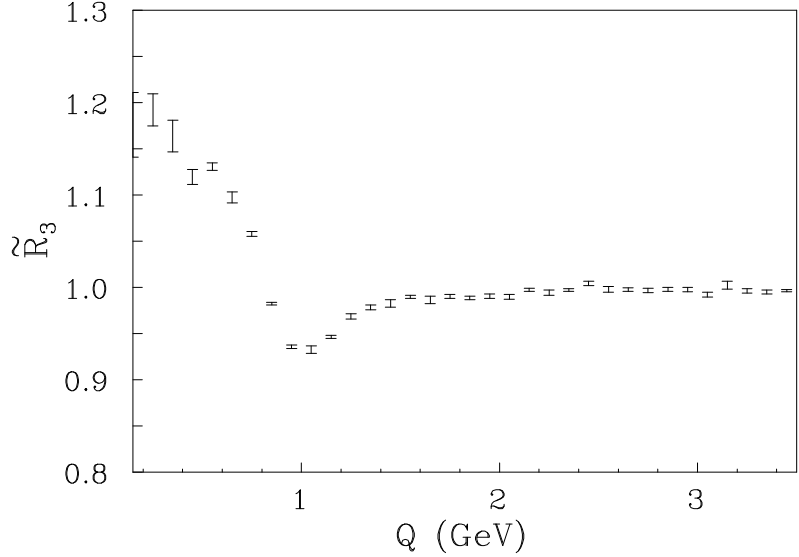,width=12cm}}
    \end{center}
  }}
\caption {\em The Q-dependence of the genuine three-particle correlation 
          function $\tilde{R}_3$, defined in the text. }
\label{f:Rtilde3}
\end{figure}

Fig~\ref{f:Rtilde3} shows $\tilde{R}_3$, the genuine three-particle correlation 
function for like-sign triplets, as approximated in Eq~(\ref{e:Rtilde3appr}). 
A strong correlation is observed for small $Q$-values. There is a dip in 
the curve for $Q$-values around $1~GeV$ which is compatible with the depletion
of $\rho^0$'s around its mass and gives an indication of the error from
using unlike-charged pions in the approximation of $\tilde{R}_3$.

\subsection*{Acknowledgements} \vspace{-2ex}
We thank T. Sj\"ostrand for very valuable discussions 
and B. S\"oderberg for discussions about permutations.



\begin{thebibliography}{99} \vspace{-1ex}
  \bibitem{hbt56}
	\bibl{R. Hanbury-Brown and R.Q. Twiss}
             {Nature}{178}{1956}{1046}
  \bibitem{mgb85}
        \bibl{M.G. Bowler}
	     {Z Phys.}{C29}{1985}{617}
  \bibitem{herwig59}
	\bibl{HERWIG 5.9; G. Marchesini, B.R. Webber, G. Abbiendi, I.G. Knowles, M.H. Seymour and L. Stanco}{Comp. Phys. Comm.}{67}{1992}{465}
  \bibitem{ts94}	
        \bibl{T. Sj\"ostrand}{Comp. Phys. Comm.}{82}{1994}{74} 
 \bibitem{ls95}
        \bibl{L. L\"onnblad and T. Sj\"ostrand}{Phys. Lett.}{B351}{1995}{293}
  \bibitem{ah86}
        \bibl{B. Andersson and W. Hofmann}
             {Phys. Lett.}{B169}{1986}{364}
  \bibitem{ba83}
        \bibl{B. Andersson, G. Gustafson, G. Ingelman and T. Sj\"ostrand}
             {Phys. Rep.}{97}{1983}{31}
  \bibitem{ags83}
        \bibl{B. Andersson, G. Gustafson and B. S\"oderberg}
             {Z. Phys.}{C20}{1983}{317}
  \bibitem{gm83}
        \bibl{N.K. Glendenning and T. Matsui}
             {Phys. Rev.}{D28}{1983}{2890}
  \bibitem{js51}
        \bibl{J. Schwinger}
             {Phys. Rev.}{82}{1951}{664}
  \bibitem{kramerskronig}
        \bibl{R. Kronig}{J. Amer. Optical Soc.}{12}{1926}{547} \\
     H.A. Kramers, {\it Atti del Congress Internationale de Fisici Como} (1927) 
  \bibitem{cs91}
	\bibl{B. Andersson, G. Gustafson, A. Nilsson and C. Sj\"ogren}
             {Z. Phys}{C49}{1991}{79}
  \bibitem{st97}
        S. Todorova, {\it Private communications}
  \bibitem{mgb86}
        \bibl{M.G. Bowler}
             {Phys. Lett}{B180}{1986}{299}
  \bibitem{aleph92}
        \bibl{D. Decamp et al. (ALEPH Coll.)}{Z. Phys.}{C54}{1992}{75}
  \bibitem{delphi94}
        \bibl{P. Abreu et al. (DELPHI Coll.)}{Z. Phys.}{C63}{1994}{17}
  \bibitem{opal91}
        \bibl{P.D. Acton et al. (OPAL Coll.)}{Phys. Lett.}{B267}{1991}{143}
  \bibitem{DELPHItuning95}
        K. Hamacher and M. Weierstall, {\it DELPHI 95-80 PHYS 515} (1995)
  \bibitem{naXX}
        \bibl{T. Alber et al., (NA35 Coll.)}{Z. Phys.}{C66}{1995}{77} \\
 	\bibl{H. B{\o}ggild et al. (NA44 Coll.)}{Phys. Rev. Lett.}{74}{1995}{3340} 	
  \bibitem{opaldelphirho}
        \bibl{P.D. Acton et al. (OPAL Coll.)}{Z. Phys.}{C56}{1992}{521} \\
        \bibl{P. Abreu et al. (DELPHI Coll.)}{Z. Phys.}{C65}{1995}{587}
  \bibitem{delphirho94}
        \bibl{P. Abreu et al. (DELPHI Coll.)}{Z. Phys.}{C63}{1994}{17}
  \bibitem{gl93}
        \bibl{G. Lafferty}{Z. Phys.}{C60}{1993}{659}
  \bibitem{3corr95}
        \bibl{P. Abreu et al. (DELPHI Coll.)}{Phys. Lett.}{B355}{1995}{415}
  \bibitem{ls96}
        B. L\"orstad and O.G. Smirnova, {\it Proceedings of the 7th International Workshop on Multiparticle Production 'Correlations and Fluctuations', Nijmegen, The Netherlands} (1996)
\end{thebibliography}
\end{document}